\documentclass{mem}
\usepackage{natbib}
\usepackage{balance}
\usepackage{txfonts}
\usepackage{graphicx}
\usepackage[a4paper]{hyperref}
\idline{00}{000}
\begin{document}

\title{Astrophysics in S.Co.P.E.}

\subtitle{}

\author{M. Brescia\inst{1}, S. Cavuoti\inst{2}, G. d'Angelo\inst{2}, 
R. D'Abrusco\inst{2}, C. Donalek\inst{3}\\ 
N. Deniskina\inst{2}, O. Laurino\inst{2}, Giuseppe Longo\inst{2}$^,$\inst{1}$^,$\inst{4}}
\offprints{G. Longo, \email{longo@na.infn.it}}

\institute{
INAF - Osservatorio Astronomico Capodimonte, via Moiariello 16, 80131, Napoli, Italy
\and
Department of Physical Sciences - University Federico II, Naples, Italy
\and
Department of Astronomy, California Institute of Technology, Pasadena, CA, USA
\and
INFN - Napoli Unit, via Cintia 9, 80126, Napoli, Italy
}

\authorrunning{M. Brescia \& et al.}

\titlerunning{astrophysics in S.Co.P.E.}

\abstract{S.Co.P.E. is one of the four projects funded by the Italian Government 
in order to provide Southern Italy with a distributed computing infrastructure for 
fundamental science. 
Beside being aimed at building the infrastructure, S.Co.P.E. is also actively pursuing 
research in several areas among which astrophysics and observational cosmology. 
We shortly summarize the most significant results obtained in the first two years of the 
project and related to the development of middleware and Data Mining tools for the 
Virtual Observatory.
\keywords{distributed computing, cosmology}}

\maketitle{}

\section{Introduction}
S.Co.P.E. is a general purpose GRID infrastructure of the University Federico II in Naples 
funded through the Italian National Plan (PON) by the Italian Government to support both 
fundamental research and small/medium size companies. 
The infrastructure has been conceived as a metropolitan GRID, embedding different (and in some cases 
pre-existing) and heterogeneous computing centers each with its specific vocation: high energy physics, 
astrophysics, bioinformatics, chemistry and material sciences, electric engineering, social sciences. 
Its intrinsically multi-disciplinary nature renders the S.Co.P.E. an ideal test bed for innovative 
middleware solutions and for interoperable tools and applications finely tuned on the needs of a 
distributed computing environment.
In what follows we shall shortly outline the main activities in the fields of astrophysics and observational 
cosmology and, in particular, we shall focus on: i) the ongoing efforts aimed at integrating the S.Co.P.E. GRID 
(hereafter SG) with the international Virtual Observatory (Sect.2), and 
ii) the implementation in the SG of the data mining (DM) VO-Neural package (Sect.3) which is developed in the framework 
of a collaboration with the Dept. of Astronomy at Caltech.  
In Sect. 4 we shortly outline a template scientific application and, finally, 
in Sect. 5, we outline some future developments.

\section{The VOb and the GRID}
The Virtual Observatory (VOb) is an international effort coordinated through the International 
Virtual Observatory Alliance \cite{IVOA} aimed at: i) federating and making interoperable 
all astronomical data archives produced by both ground based and space borne instruments; 
ii) deploying a new generation of science applications or tools which use VOb protocols for 
exploratory data analysis and for the extraction of knowledge from massive data sets.
The VOb is inherently distributed: data collections remain with their 
providers and are accessed through standard interfaces. 
The access to the data takes place through a registry which contains information about data sets, 
archives, catalogs, surveys, and computational services that can be accessed through VOb 
interfaces \cite{deyoung}.
While the federation and fusion of heterogeneous data archives and the implementation of 
flexible data reduction and data analysis tools have been widely addressed and, at least 
in their fundamental aspects, solved, the possibility to access large distributed computing 
facilities to perform computing intensive tasks has not yet been satisfactorily answered.

One problem to be solved is the conflict existing between the VOb and the GRID 
security procedures: most users of a specific Virtual Organization (VO) do not possess the 
personal certificates which are requested to access the GRID or, even when they do have a 
personal certificate, the computing GRID which they need does not recognize their own
certification authority. 

In the framework of the VONeural project (Sect. \ref{voneural2}) and in order to make our
Data Mining (DM) tools accessible to the wider community, we implemented and tested a general 
purpose interface between the UK-ASTROGRID \cite{astrogrid} and the SG.

\subsection{GRID-Launcher v.1.0}
The UK based ASTROGRID is one of the most robust astronomical Virtual Organizations so far
implemented and represents a good ground for testing innovative solutions. 
The main problem we had to face was the fact that most users which are recognized by 
the AG User Authentication Service do not possess a personal GRID certificate 
and cannot therefore access distributed computing resources.
This problem can be at least in part circumvented by offering the applications as
web services to be consumed on the Grid via a service certificate (or ''robot'' certificates).
At the time GRID-Launcher v.1.0 was developed, this option had not yet been formally 
accepted by the EGEE-2 boards and we were obliged to implement a test version which 
makes use of a personal GRID certificate (signed by the INFN-GRID CA) 
which is recognized by the S.Co.P.E. GRID.

\begin{figure}
\resizebox{\hsize}{!}{\includegraphics[clip=true]{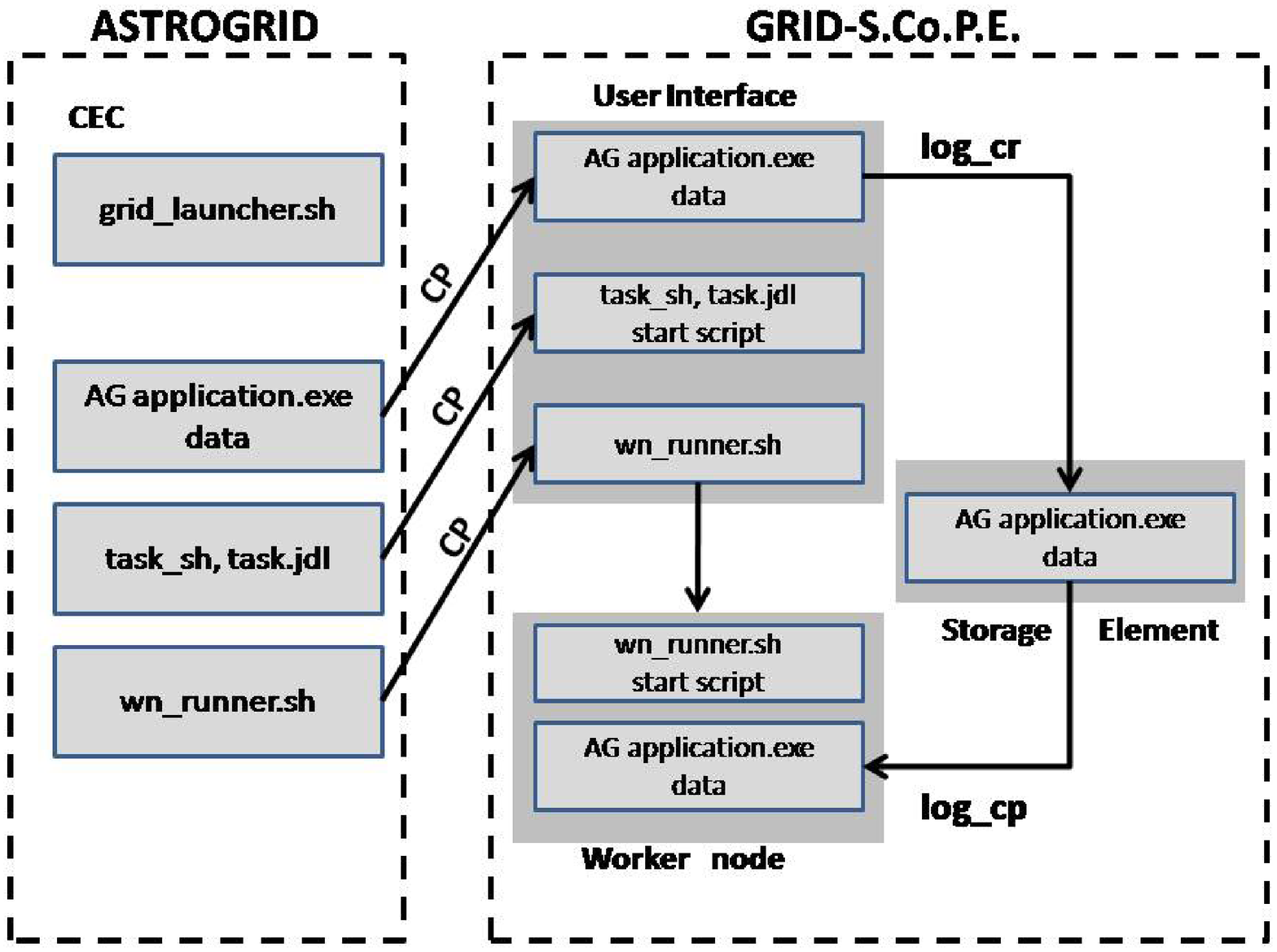}}
\resizebox{\hsize}{!}{\includegraphics[clip=true]{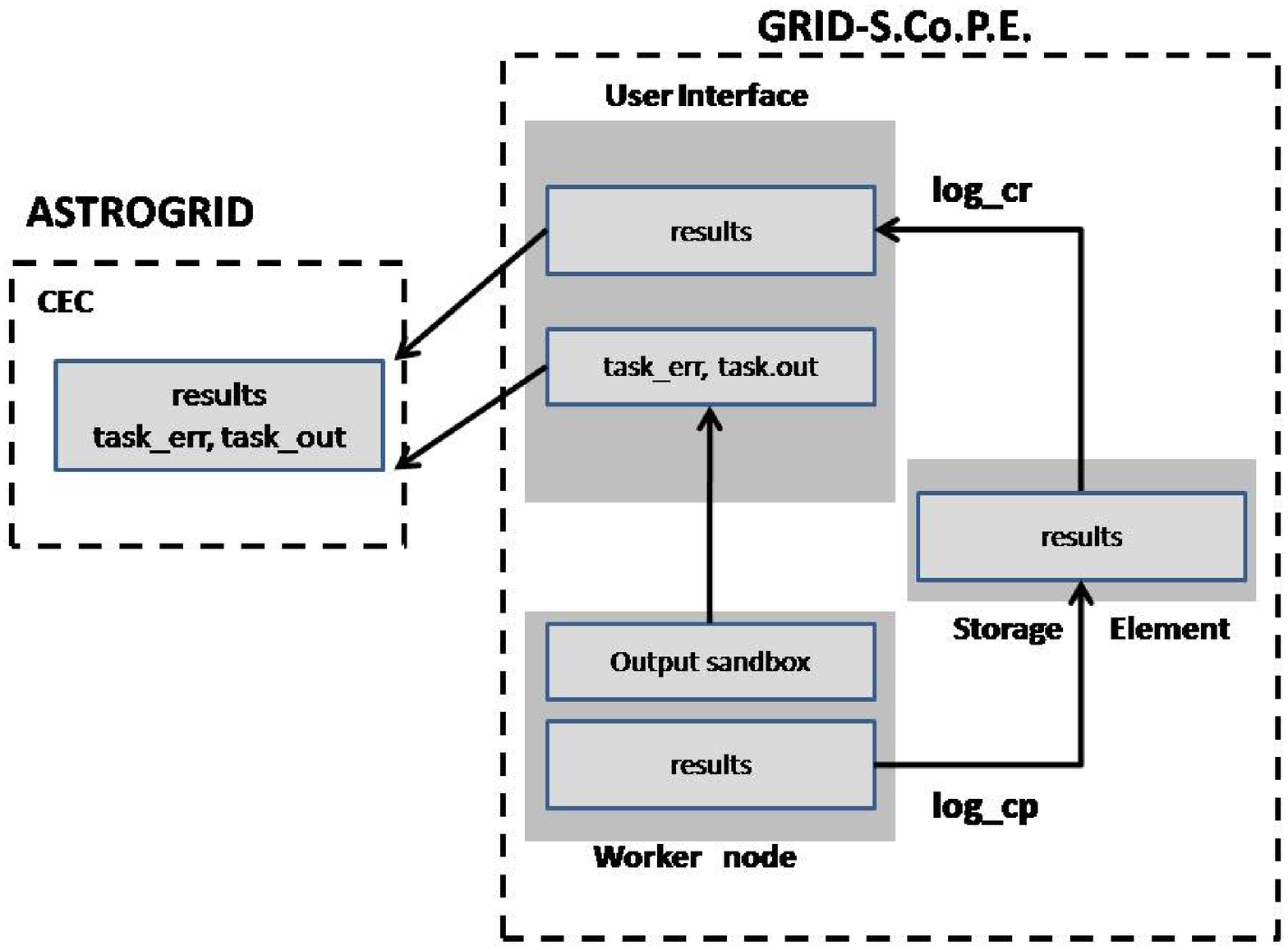}}
\caption{\footnotesize Grid Launcher v.1.0 workflows for input and output.
UI: user interface; RB: resource broker; SE: storage element; CE: computing element; 
WN: working node. Upper panel: input flow; lower panel: output flow.} \label{figa}
\end{figure}

\noindent In a very schematic way, GRID-Launcher works as it is summarized in Fig. \ref{figa}:

\begin{itemize}
\item It takes the user input from the User Interface of the ASTROGRID Desktop, collects all 
files, tabs and programs needed and generates automatically three scripts: $task.sh$, $task.jdl$ 
and $wn\_runner.sh$ to be executed on the GRID;
\item it wraps them in an archive and sends it to the GRID UI (authentication 
takes place with public keys exchange);
\item the UI unpacks them, copies the data to the Storage Element (SE), copies $wn\_runner.sh$ to the WN's, 
starts $task.sh$ and $task.jdl$;
\item $wn\_runner.sh$ starts on the WNs, takes the data from SE, starts the application and puts the results 
on the SE. The GRID generates automatically two output files $task.err$ and $task.out$ and sends them to the UI 
using the Output SandBox.
\item GRID-launcher periodically checks the status of job and when it ends, it moves the results from the UI 
to the ASTROGRID machine. GRID-launcher receives the data archive, unpacks them and puts the results into the 
AG Myspace (VO-Space).
\end{itemize}

\noindent So far, GRID-Launcher v.1.0 has been implemented and tested on an handful of applications: VO-Neural\_MLP 
\& VO-Neural\_SVM (cf. Sect. \ref{voneural2}), Sextractor \cite{SEx} \& SWarp \cite{swarp}.

\section{VO-Neural}\label{voneural2}
As it was mentioned above, in the last decade many national \cite{nvo} and international \cite{EuroVO} projects have 
solved many problems related to the federation of heterogeneous data sets while much remains still to be done for 
what tools and user interfaces are concerned. 
One of the main issues to be solved is the implementation of scalable and user friendly data mining tools capable to 
deal with the huge VOb data sets. 

VO-Neural is a data mining framework capable to work on massive ($>1$ TB) data sets (catalogues) in a 
distributed computing environment matching the IVOA standards and requirements. 
VO-Neural is the evolution of the AstroNeural \cite{astroneural} project which was started in 1994, as 
a collaboration between the Department of Mathematics and Applications at the University of Salerno and 
the Astronomical Observatory of Capodimonte-INAF, and is currently under continuous evolution.
VO-Neural allows to extract from large datasets information useful to determine patterns, relationships, 
similarities and regularities in the space of parameters, and to identify outlayers. 
In its final version, it will be accessible both as a web application and through the AG Desktop and
will offer main elaborative features like exploratory data analysis, data prediction 
and ancillary functionality like fine tuning, visual exploration of the main characteristics of the datasets, 
etc..
Besides offering the possibility to use the individual routines to perform specific tasks, VO-Neural will
provide the user with a complete framework to write his own customized programs.
In the next two paragraphs we shortly outline the main features of two supervised clustering
models already included in the package which have already been used on the GRID-S.Co.P.E. for specific 
science applications.

\subsubsection{$VONeural\_MLP$}
$VONeural\_MLP$ is an implementation of a standard Multi Layer Perceptron based on the FANN 
(Fast Artificial Neural Networks) Library \cite{fann}, written in C++ \cite{skordovski}.
The algorithm known as Multi Layer Perceptron (MLP) is based on the concept of perceptron
and the method of learning is based on gradient-descent method that allows to find a local minimum 
of a function in a space with N dimensions. 
The weights associated to the connections between the layers of neurons are initialized 
at small and random values, and then the MLP applies the learning rule using part of 
the template patterns.
Once convergence has been achieved and a validation procedure has been applied in order to 
avoid overfitting, the performances of the network are evaluated on a disjoint test set extracted from
the template patterns. The resulting network is then applied to the original data.

\subsubsection{$VONeural\_SVM$}
$VONeural\_SVM$ is an implementation of the Support Vector Machines \cite{russo,cavuoti} based on
the LIBSVM library \cite{libsvm}. 
Support Vector Machines perform classification of records into classes by first mapping the data into 
an higher dimensionality and then using a set of template vectors (targets) to find in this new space 
a separation hyperplane with the largest margin.
Without entering into details \cite{boser}, we shall just remember that, 
in the case of the C-SVC implemented with the RBF (Radial Basis Functions) kernel, the position of this hyperplane depends on two 
parameters ($C$ and $\gamma$) which cannot be estimated in advance but need to be evaluated by finding the maximum in a grid of 
values which is usually defined by letting $C$ and $\gamma$ vary as $C = 2^{-5}, 2^{-3}, ..., 2^{15}$ and $\gamma = 2^{-15}, 2^{-13}, ..., 2^3$.
Due to their computational weight, and to the need to run many iterations for different pairs of the two parameters,
SVM are ideally suited for being used on the GRID.

\section{An application to the classification of AGN in the SDSS}
The astronomical community is used to perform DM tasks in a sort of ''hidden'' way (cf. the case of specific objects 
selection in a color-color diagram) but it has not yet become familiar with the potentialities of more advanced tools 
such as those described here.
This is mainly due to the fact that these tools are often everything but user friendly and require an in depth 
understanding of the (often complex) theory laying behind them; a complexity which often discourages potential users.
Therefore, a crucial aspect of the project is the application to challenging problems capable to exemplify the new 
science which will emerge from the adoption of a less conservative approach to the analysis of the data.
Two science cases, namely the evaluation of photometric redshifts (a regression and classification problem
based on the use of $VONeural\_MLP$) and the selection of candidate quasars in the Sloan Digital Sky Survey \cite{SDSS}
(based on the use of unsupervised clustering algorithms and agglomerative clustering) have already been published 
in the literature \cite{dabrusco,dabrusco_1}. 
We shall therefore focus on the application of $VONeural\_SVM$ to the classification of AGNs.

\noindent The classification of AGN is usually performed on their overall spectral distribution using some 
spectroscopic indicators (equivalent linewidths, FWHM of specific lines or lines flux ratios) and diagnostic 
diagrams (usually called BPT) which are difficult and time consuming to derive.
In this diagrams AGN and not-AGN are empirically separated by some lines derived either from the theory or from
empirical laws such as those derived by \cite{kewley,kauffman,heckman}.
A reliable and accurate AGN classificator based on photometric features only, would allow to save precious telescope 
time and enable several studies based on statistically significant samples of objects. 
We therefore used a supervised clustering of the photometric data exploiting the information contained 
in a spectroscopic Base of Knowledge (BoK) derived from available catalogues. 
We wish to stress that since neural networks have no power of extrapolation all the biases in the BoK are 
reproduced and therefore the BoK needs to be as complete and bias-free as possible.
As classification tools, we used both the MLP and, due to the intrinsically binary nature of the problem (AGN against non-AGN, 
Seyfert 1 against Seyfert 2, etc) also the SVM. 
\begin{table*}\label{results}
\begin{tabular}{llcrr}
\hline
\hline
experiment       & BoK                    & algorithm & efficiency & completeness\\
\hline
AGN vs Mix       & BPT plot + Kewley line & MLP       & $76\%$     & $54\%$      \\
                 & BPT plot + Kewley line & SVM       & $74\%$     & $55\%$      \\
\hline
Type 1 vs 2      & BPT plot + Kewley line & MLP       & $95\%$     & $\sim 100\%$\\
                 & BPT plot + Kewley line & SVM       & $82\%$     & $98\%$      \\
\hline
Seyfert vs LINER & BPT plot + Hecman \& Kewley lines & MLP       & $80\%$     & $92\%$      \\
                 & BPT plot + Kewley line & SVM       & $78\%$     & $89\%$      \\
\hline
\end{tabular}
\caption{Summary of the results of supervised classification experiments performed using both 
$VONeural\_MLP$ and $VONeural\_SVM$.}
\end{table*}

\noindent The BoK was obtained from the fusion of two catalogues.
\begin{itemize}
\item \cite{sorrentino} separated objects into Seyfert 1, Seyfert 2 and ''Not AGN'' 
using the Kewley's lines \cite{kewley}; 
\item a catalogue derived by us from the SDSS spectroscopic archive using the criteria introduced by \cite{kauffman} 
in which objects are classified as AGN, not AGN, and ''mixed''. 
The Mix and Pure AGN zone were further divided into Seyfert and LINERs by using the Heckman line\cite{heckman}.
\end{itemize}
\noindent We made three experiments using both the MLP and SVM, and for all of them we used the same set of features 
(for a definition refer to the SDSS specifications) extracted from the SDSS database: 
$petroR50\_u$, $petroR50\_g$, $petroR50\_r$, $petroR50\_i$, $petroR50\_z$, $concentration\_index\_r$, $fibermag\_r$, 
$(u-g) dered$, $(g-r) dered$, $(r-i) dered$, $(i-z) dered$, $dered\_r$, together with the photometric redshift 
in \cite{dabrusco_1}.
We performed three types of classification experiments: AGN vs Mix, Type1 vs Type2, Seyfert vs LINER. 
The experiments with SVM were performed on the SG using 110 worker nodes. In order to test the interoperability of
the four PON projects, the 110 nodes were taken from the Napoli, Catania and Cagliari PON locations.
Results are summarized in Table \ref{results} and, as it can be seen, the use of machine learning tools allows 
to reach performances which in some cases (e.g. Type 1 vs 2 with MLP's) cannot by any means be achieved with more 
traditional tools. 
A more detailed discussion of the results will be presented in (Cavuoti, d'Abrusco \& Longo, 2008, in preparation).

\section{Future developments}
We plan to continue the development of VO-Neural and to offer it as a web application. 
More in detail, we plan to deploy a general purpose GRID-Launcher interface capable to 
launch any ''command line'' program through a ''robot certificate'' (GRID-Launcher 2.0).

At the moment we are engineering the package in order to increase its flexibility and capability 
to adapt to a distributed computing environment. 
We are also implementing parallel versions of some tools which are particularly demanding in terms 
of computing time.
We also plan to integrate, within the VO-Neural interface existing statistical software 
(such as, for instance, the VO-STAT web application \cite{vostat}), in order to ensure proper statistical 
tools for exploratory data analysis and for the evaluation of the results.
The status of the project can be monitored at the URL: http://voneural.na.infn.it/ .
\\

\noindent {\it Acknowledgements.} 
The authors wish to thank M. Paolillo and E. de Filippis for many useful discussions. 
The work was funded through the Euro VO-Tech project and the MUR funded PON-S.Co.P.E. 

\bibliographystyle{aa}

\begin{thebibliography}{}
\bibitem[Bertin \& Arnouts 1996]{SEx} Bertin E. \& Arnouts S. 1996, A\& AS, 117, 393.
\bibitem[Boser et al. 1992]{boser} Boser B. E., Guyon I. \& Vapnik V. 1992,in Proc. of the Fifth Annual Workshop on
Computational Learning Theory, 144, ACM Press. 
\bibitem[Cavuoti 2008]{cavuoti} Cavuoti S. 2008, M.Sc. Thesis, University of Napoli Federico II.
\bibitem[D'Abrusco et al. 2007]{dabrusco} D'Abrusco R., Staiano A., Longo G., Brescia M., De Filippis E., Paolillo M., 
Tagliaferri R. 2007, ApJ, 663, 752.
\bibitem[D'Abrusco et al. 2008]{dabrusco_1} D'Abrusco R., Longo G. \& Walton N.A. 2008, astro-ph/0805.0156v1.
\bibitem[Hanish \& De Young 2008]{deyoung} Hanish R. \& de Young D. 2008, in The National Virtual Observatory Book
ASP Conference Series, Vol. 382, M. J. Graham M. J. Fitzpatrick, \& T. A. McGlynn eds., 1.
\bibitem[Heckman 1980]{heckman} Heckman T.M. 1980, A\& A, 87, 182.
\bibitem[Kauffman et al. 2003]{kauffman}Kauffman G. et al. 2003, MNRAS, 346, 1055.
\bibitem[Kewley et al. 2006]{kewley} Kewley L.J. et al. 2006, MNRAS, 372, 961.
\bibitem[Skordovski 2008]{skordovski} Skordovski B. 2008, M.Sc. Thesis, University of Napoli Federico II.
\bibitem[Russo 2007]{russo} Russo V. 2007, M.Sc. Thesis, University of Napoli Federico II.
\bibitem[cf. Stoughton et al. 2002]{SDSS} Stoughton C., Lupton R. H., Bernardi M. et al. 2002, AJ, 123, 485.
\bibitem[Tagliaferri et al. 2003]{astroneural} Tagliaferri R., Longo G., Milano L., Acernese F., et al. 2003, 
Neural Networks, 16, 297.
\bibitem[Sorrentino et al 2006]{sorrentino} Sorrentino G., Radovich M. \& Rifatto A. 2006, A \& A 451, 809.
\bibitem[IVOA; URL.1]{IVOA} URL.1: IVOA: http://www.ivoa.org/
\bibitem[cf. URL.2]{nvo} URL.2: NVO: http://www.nvo.org/
\bibitem[cf. URL.3]{EuroVO} URL.3: Euro-VO: http://www.eurovo.org/
\bibitem[hereafter AG; URL.4]{astrogrid} URL.4: http://www.astrogrid.uk/
\bibitem[URL.5]{voneural} URL.5: http://voneural.na.infn.it/
\bibitem[URL.6]{fann} URL.6: http://leenissen.dk/fann/
\bibitem[URL.7]{libsvm} URL.7: http://www.csie.ntu.edu.tw/~cjlin/libsvm/
\bibitem[URL.8]{swarp} URL.8: SWarp User manual , E. Bertin at: http://terapix.iap.fr/rubrique.php?id\_rubrique=49
\bibitem[URL.9]{vostat} URL.9: VO-STAT, http://astrostatistics.psu. edu/
\end{thebibliography}

\end{document}